\documentclass[11pt,preprint]{aastex}
\usepackage{emulateapj5}

\newcommand{\cf}{{\ifmmode{C_f}\else{$C_{f}$}\fi}}
\newcommand{\zem}{{\ifmmode{z_{em}}\else{$z_{em}$}\fi}}
\newcommand{\zabs}{{\ifmmode{z_{abs}}\else{$z_{abs}$}\fi}}
\newcommand{\kms}{{\ifmmode{{\rm km~s}^{-1}}\else{km~s$^{-1}$}\fi}}
\newcommand{\vej}{{\ifmmode{v_{ej}}\else{$v_{ej}$}\fi}}
\newcommand{\cmmm}{{\ifmmode{{\rm cm}^{-3}}\else{cm$^{-3}$}\fi}}
\newcommand{\lya}{Ly$\alpha$}

\newcounter{species} 
\def\ion#1#2{\setcounter{species}{#2}#1$\;${\scriptsize\Roman{species}}\relax}

\slugcomment{\today} 

\shorttitle{Monitoring Outflow Winds in SDSS~J1029+2623}
\shortauthors{Misawa et al.}

\begin{document}

\title{Resolving the Clumpy Structure of the Outflow Winds in the
  Gravitationally Lensed Quasar SDSS~J1029+2623\altaffilmark{1}}

\footnotetext[1]{Based on data collected at Subaru Telescope, which is
  operated by the National Astronomical Observatory of Japan.}

\author{Toru Misawa\altaffilmark{2},
        Naohisa Inada\altaffilmark{3},
        Masamune Oguri\altaffilmark{4,5,6},
        Poshak Gandhi\altaffilmark{7,8},
        Takashi Horiuchi\altaffilmark{9},
        Suzuka Koyamada\altaffilmark{9},
        Rina Okamoto\altaffilmark{9}}

\altaffiltext{2}{School of General Education, Shinshu University,
  3-1-1 Asahi, Matsumoto, Nagano 390-8621, Japan}
\altaffiltext{3}{Department of Physics, Nara National College of
  Technology, Yamatokohriyama, Nara 639-1080, Japan}
\altaffiltext{4}{Research Center for the Early Universe, University of
  Tokyo, 7-3-1 Hongo, Bunkyo-ku, Tokyo 113-0033, Japan}
\altaffiltext{5}{Department of Physics, University of Tokyo, 7-3-1
  Hongo, Bunkyo-ku, Tokyo 113-0033, Japan}
\altaffiltext{6}{Kavli Institute for the Physics and Mathematics of
  the Universe (Kavli IPMU, WPI), University of Tokyo, Chiba 277-8583,
  Japan}
\altaffiltext{7}{Department of Physics, Durham University, South Road,
  Durham DH1 3LE, UK}
\altaffiltext{8}{Institute of Space and Astronautical Science (ISAS),
  Japan Aerospace Exploration Agency, 3-1-1 Yoshinodai, chuo-ku,
  Sagamihara, Kanagawa 252-5210, Japan}
\altaffiltext{9}{Department of Physics, Faculty of Science, Shinshu
  University, 3-1-1 Asahi, Matsumoto, Nagano 390-8621, Japan}
\email{misawatr@shinshu-u.ac.jp}

\begin{abstract}
We study the geometry and the internal structure of the outflowing
wind from the accretion disk of a quasar by observing multiple
sightlines with the aid of strong gravitational lensing. Using
Subaru/HDS, we performed high-resolution ($R$ $\sim$ 36,000)
spectroscopic observations of images~A and B of the gravitationally
lensed quasar SDSS~J1029+2623 (at \zem\ $\sim$ 2.197) whose image
separation angle, $\theta$ $\sim$ 22$^{\prime\prime}\!\!$.5, is the
largest among those discovered so far.  We confirm that the difference
in absorption profiles in the images~A and B discovered by
\citet{mis13} remains unchanged since 2010, implying the difference is
not due to time variability of the absorption profiles over the delay
between the images, $\Delta t$ $\sim$ 744~days, but rather due to
differences along the sightlines. We also discovered time variation of
\ion{C}{4} absorption strength in both images~A and B, due to change
of ionization condition. If a typical absorber's size is smaller than
its distance from the flux source by more than five orders of
magnitude, it should be possible to detect sightline variations among
images of other smaller separation, galaxy-scale gravitationally
lensed quasars.
\end{abstract}


\section{INTRODUCTION \label{sec1}}
Outflowing winds from accretion disks, accelerated by radiation force
\citep{mur95,pro00}, magnetocentrifugal force \citep{eve05,dek95},
and/or thermal pressure \citep{bal93,kro01,che05}, are a key
evolutionary link between quasars and their host galaxies. The disk
outflow plays an important role as 1) it extracts angular momentum
from the accretion disk, leading to growth of black holes
\citep{bla82,kon94,eve05}, 2) it provides energy and momentum feedback
and inhibits star formation activity \citep[e.g.,][]{spr05}, and 3) it
induces metal enrichment of intergalactic medium (IGM)
\citep{ham97b,gab06}.  Outflowing matter has been detected through
blueshifted absorption lines in $\sim$70\%\ of quasar spectra
\citep{ham12}.  These are usually classified as {\it intrinsic}
absorption lines, and distinguished from {\it intervening} absorption
lines that originate in foreground galaxies or in the IGM. Thus,
intrinsic absorption lines are powerful and unique tool to probe the
outflows in quasars.  However, the main challenge in their study is
that these are traceable along {\it only} single sight-lines (i.e., a
one dimensional view alone) toward the nucleus for each quasar,
whereas the absorber's physical conditions likely depend strongly on
polar angle \citep[e.g.,][]{gan01,elv00}.

Multiple images of quasars produced by the gravitational lensing
effect provide a unique pathway for studying the multiple sightlines,
a technique frequently applied for intervening absorbers (e.g.,
\citealt{cro98,rau99,lop05}).  It is clear that lensed quasars with
larger image separation angles have more chance of detecting
structural differences in the outflow winds in the vicinity of the
quasars themselves.  In this sense, the following three lensed quasars
are the most promising site for our study as they are lensed by a
cluster of galaxies rather than a single massive galaxy:
SDSS~J1004+4112 with separation angle of $\theta$ $\sim$
14$^{\prime\prime}\!\!$.6 \citep{ina03}, SDSS~J1029+2623 with $\theta$
$\sim$ 22$^{\prime\prime}\!\!$.5 \citep{ina06,ogu08}, and
SDSS~J2222+2745 with $\theta$ $\sim$ 15$^{\prime\prime}\!\!$.1
\citep{dah13}.  \citet{gre06} proposed that the differences in
emission line profiles between the lensed images of SDSS~J1004+4112
can be explained by differential absorptions along each sight-line
although no absorption features are detected.

There are clear absorption features detected at the blue wings of the
\ion{C}{4}, \ion{N}{5}, and \lya\ emission lines of other lensed
quasar, SDSS~J1029+2623 at \zem\ $\sim$ 2.197\footnote[10]{The quasar
  redshift was derived from \ion{Mg}{2} emission lines in
  \citet{ina06}.  An uncertainty of \zem\ is $\Delta z$ $\sim$ 0.0003,
  corresponding to $\Delta v$ $\sim$ 30~\kms. The \zem\ could be
  blueshifted from the systemic redshift by $\sim$100~\kms\ in average
  \citep{tyt92}.}, the current record-holding large-separation quasar
lens, in low/medium resolution spectra \citep{ina06,ogu08}.
\citet{mis13} obtained high-resolution spectra of the brightest two
images (i.e., images~A and B), and carefully deblended the \ion{C}{4}
and \ion{N}{5} absorption lines into multiple narrower components.
They show several clear signs supporting an origin in the outflowing
wind rather than in foreground galaxies or the IGM \citep{mis13}: i)
partial coverage, i.e., the absorbers do not cover the background flux
source completely along our sightline, ii) line-locking, iii) large
velocity distribution (FWHM $\geq$1000~km/s), and iv) a small ejection
velocity\footnote[11]{The ejection velocity is defined as positive if
  the absorption line is blueshifted from the quasar emission
  redshift.} from the quasar.  \citet{mis13} also discovered a clear
difference in parts of these lines between the images~A and B in all
\ion{C}{4}, \ion{N}{5}, \lya\ absorption lines, which can be explained
by the following two scenarios: (a) intrinsic time variability of the
absorption features over the time delay of the two images
\citep[e.g.,][]{cha07}, and (b) a difference in the absorption levels
between the different sight-lines of the outflowing wind
\citep[e.g.,][]{che03,gre06}.  With a single epoch observation, we
cannot distinguish these scenarios.

In this letter, we present results from new spectroscopic observations
of SDSS~J1029+2623 conducted $\sim$4 years later with the goal of
conclusively determining the origin of the difference in the
absorption features.  We also examine the global and internal
structure of the outflow.  Observations and data reduction are
described in \S2, and results and discussion are in \S3 and \S4.

\section{OBSERVATIONS and DATA REDUCTION \label{sec2}}
We observed the images~A and B of SDSS~J1029+2623 with Subaru/HDS on
2014 April 4 (the 2014 data, hereafter), 1514~days after the previous
observation on February 2010 (the 2010 data; \citealt{mis13}).  The
time interval between the observations is longer than the time delay
between images~A and B, $\Delta t$ $\sim$ 744~days in the sense of A
leading B \citep{foh13}.  We have taken high-resolution spectra ($R$
$\sim$ 36,000) with a slit width of $1.\!\!^{\prime\prime}0$, while
\citet{mis13} took $R$ $\sim$ 30,000 spectra using
$1.\!\!^{\prime\prime}2$ slit width.  The wavelength coverage is
3400--4230 \AA\ on the blue CCD and 4280--5100~\AA\ on the red CCD,
which covers \lya, \ion{N}{5}, \ion{Si}{4}, and \ion{C}{4} absorption
lines at \zabs\ $\sim$ \zem.  We also sampled every 4 pixels in both
spatial and dispersion directions (i.e., $\sim$0.05\AA\ per pixel) to
increase S/N ratio.  The total integration time is 11400~s and the
final S/N ratio is about 14~pix$^{-1}$ for both of the images.

We reduced the data in a standard manner with the software
IRAF\footnote[12]{IRAF is distributed by the National Optical
  Astronomy Observatories, which are operated by the Association of
  Universities for Research in Astronomy, Inc., under cooperative
  agreement with the National Science Foundation.}.  Wavelength
calibration was performed using a Th-Ar lamp.  We applied flux
calibrations for quasar spectra using the spectrophotometric standard
star Feige34\footnote[13]{We reduced the 2010 data again by applying
  flux calibration, while \citet{mis13} only presented normalized
  spectra.  Because the continuum fitting gives an additional
  uncertainty for absorption depth and profile we use flux-calibrated
  spectra in this study.}.  We did not adjust a spectral resolution of
the 2014 spectra ($R$ $\sim$ 36,000) to the 2010 ones ($R$ $\sim$
30,000) before comparing them, because a typical line width of each
absorption component after deblending into multiple narrower
components is large enough (FWHM $\geq$ 10~\kms; \citealt{mis13}) to
ignore the influence of spectral resolution.

\section{RESULTS \label{sec3}}
Here, we examine the time variability of \lya, \ion{N}{5}, and
\ion{C}{4} lines.  Although \ion{Si}{4} is also detected, we cannot
use it for the analysis because the \ion{Si}{4} line is severely
contaminated by intervening \ion{Si}{2} and \ion{C}{4} lines at lower
redshift.  For the purpose of comparing absorption profiles, we
increase the S/N ratio by resampling the spectra every 0.5\AA. For a
more quantitative test, we also compare the flux difference between
the two spectra to 3$\sigma$ flux uncertainty\footnote[14]{Total flux
  uncertainty is calculated by $\sigma$ = $\sqrt{\sigma_1^2 +
    \sigma_2^2}$, where $\sigma_1$ and $\sigma_2$ are the one sigma
  errors of the first and second spectra, respectively, and include
  photon-noise and readout-noise.}.

\begin{figure}
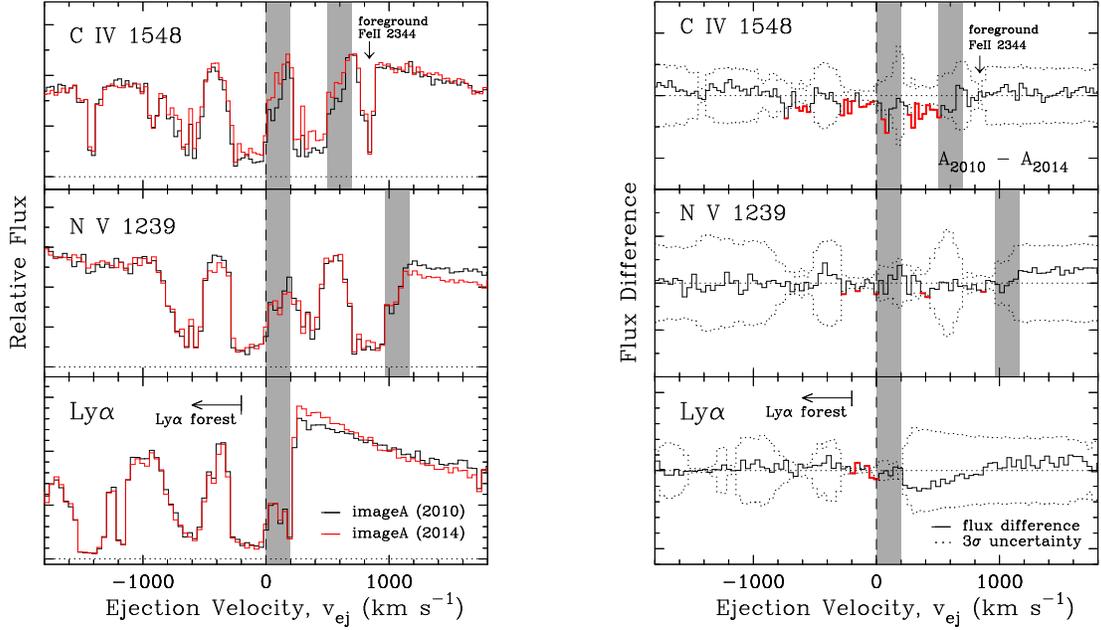

 \begin{center}
  \includegraphics[width=8cm,angle=0]{fig1a.eps}
  \includegraphics[width=8cm,angle=0]{fig1b.eps}
 \end{center}
 \caption{Comparison of the \ion{C}{4}, \ion{N}{5}, and
   \lya\ absorption profiles taken in 2010 (black) and 2014 (red)
   toward the image~A of SDSS~J1029+2623.  (Left) The horizontal axis
   denotes ejection velocity relative to the quasar emission redshift
   (\zem\ $\sim$ 2.197).  The vertical axis flux scale is arbitrary.
   The shaded regions cover the $C_1$ of \lya\ line and the doublets
   of \ion{C}{4} and \ion{N}{5} lines while the other regions are the
   $C_2$. Intervening absorption lines that are not related to the
   outflow are marked with arrows and transition names.  (Right)
   Comparison of flux difference between two spectra (solid histogram)
   with flux uncertainty (dotted histogram). Spectral regions where
   absorption line show variation in more than 3$\sigma$ level are
   marked with red histogram, except for \lya\ forest and intervening
   absorption lines.\label{f1}}
\end{figure}

As shown in Figure~\ref{f1}, we did not find any time variations
either in \lya\ or \ion{N}{5} absorption lines. But \ion{C}{4} lines
in both of the images~A and B showed clear variation in its line {\it
  strength} by more than the 3$\sigma$ level, without any change in
line {\it profiles}.

\begin{figure}
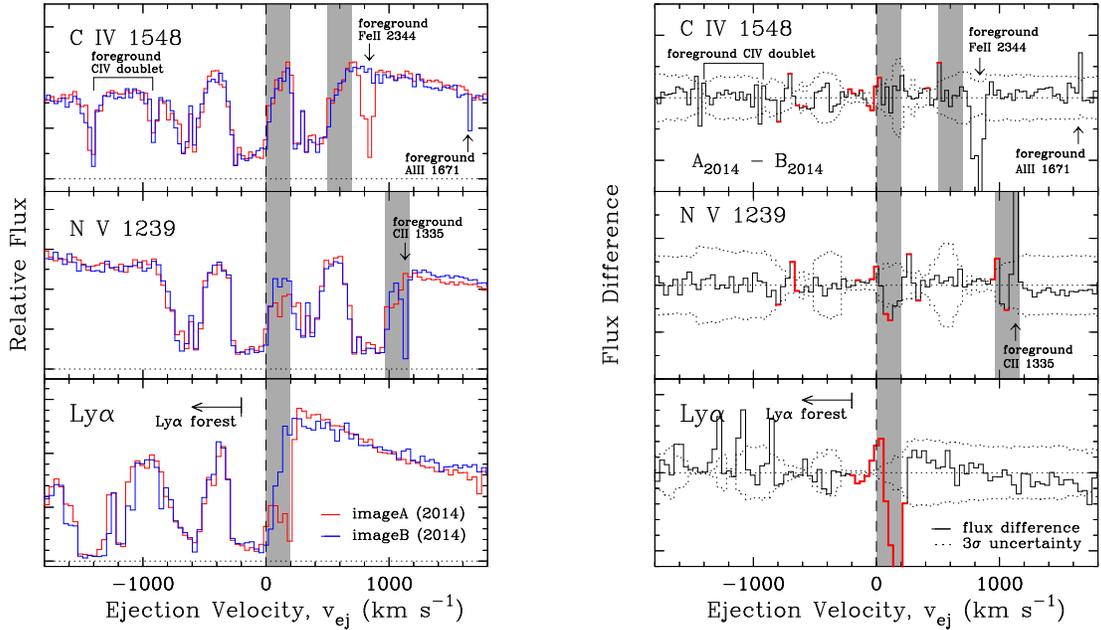

 \begin{center}
  \includegraphics[width=8cm,angle=0]{fig2a.eps}
  \includegraphics[width=8cm,angle=0]{fig2b.eps}
 \end{center}
 \caption{Same as Figure~\ref{f1} but for the images~A (red) and B
   (blue) of the 2014 spectra.  The \ion{C}{4} doublet at \vej\ $\sim$
   $-$1400~\kms, which shows $\geq$ 3$\sigma$ difference, is probably
   an intervening absorption line because the profile is narrow and
   the flux at the line center approaches close to zero before
   sampling.  The profile difference in the shaded regions between the
   images~A and B detected in the 2010 data \citep{mis13} still
   remains in the 2014 data.\label{f2}}
\end{figure}

In the 2010 spectra, \citet{mis13} discovered a clear difference in
parts of \ion{C}{4}, \ion{N}{5}, and \lya\ absorption lines at an
ejection velocity of \vej\ $\sim$ 0 -- 200~\kms\ between the images~A
and B.  The difference still remains at $\geq$ 3$\sigma$ level in the
2014 spectra, except for the \ion{C}{4} absorption lines for which the
difference between images~A and B is no longer significant
(Figure~\ref{f2}).  Thus, absorption components at \vej\ $\sim$ 0 --
200~\kms\ (shaded regions in Figures~\ref{f1} and \ref{f2}) probably
have a different origin from the other absorption components, as
suggested in \citet{mis13}.  We call the former and the latter
  components as Components 1 and 2 ($C_1$ and $C_2$, hereafter),
  respectively, and distinguish them in the discussion below.

\section{DISCUSSION \label{sec4}}
In this section, we discuss the difference between the images~A and B,
the origin of the time variation seen in the \ion{C}{4} lines between
the 2010 and the 2014 data, and then the detectability of the
sightline variation for quasar images with smaller separations, lensed
by a single galaxy.

\subsection{Difference between the Images~A and B}
\citet{mis13} presented two plausible scenarios that explain the
sightline variation of the $C_1$: (a) time variability over the time
delay between the images, $\Delta t$ $\sim$ 744~days, and (b) the
difference in the absorption levels between two sightlines. With our
new data, we can reject the first scenario because $C_1$ is again
detected only in the image~A as in the 2010 data.  If this is due to
time variation, the $C_1$ has to decrease (from the image~A to B in
2010), increase (from the image~B in 2010 to the image~A in 2014), and
then decrease again (from the image~A to B in 2014), which requires an
unlikely fine-tuning.  Although sightline variations are often
observed in intervening absorption lines
\citep[e.g.,][]{cro98,rau99,lop05}, the $C_1$ in the image~A should
have its origin in {\it intrinsic} absorber because it shows partial
coverage \citep{mis13} and time variation (this study).  Thus, the
geometry of the outflow is such that the $C_1$ covers only sightline
to the image~A, while the $C_2$ covers both sightlines of the images~A
and B.

\begin{figure}
 \begin{center}
  \includegraphics[width=15cm,angle=0]{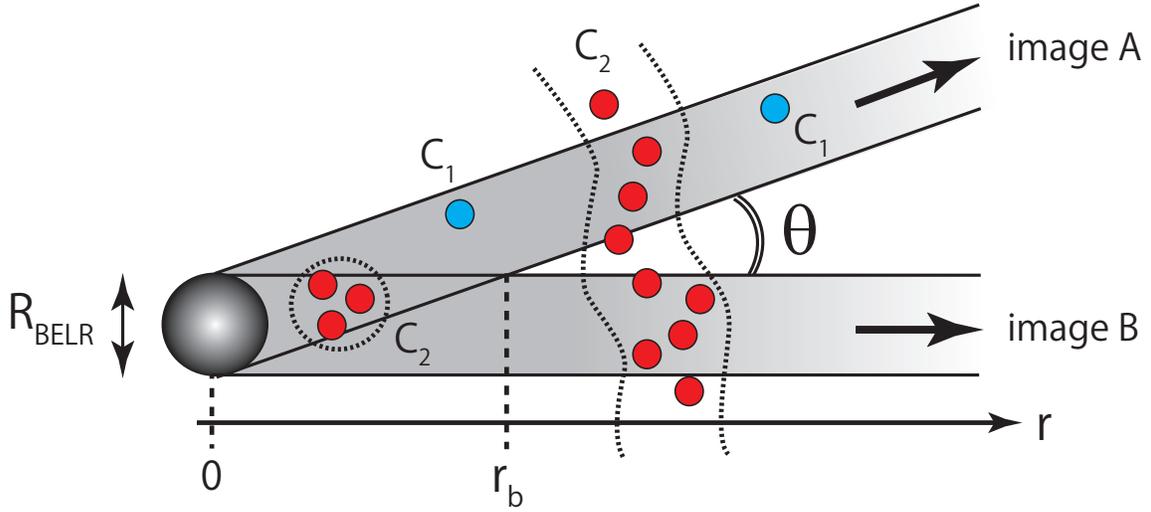}
 \end{center}
 \caption{Possible locations of the $C_1$ and $C_2$ absorbers toward
   our sightlines.  The large sphere represents the broad emission
   line region as a background flux source, while small blue and red
   filled circles are the $C_1$ and $C_2$ absorbers.  The $C_1$
   absorber always locates on only the sightline toward the image~A,
   regardless of its distance.  $C_2$ locates on both sightlines
   toward the images~A and B or has a filamentary (or sheet-like)
   structure consisting of number of small clumpy clouds.  The
   boundary distance ($r_b$) is the distance at which the two
   sightlines become fully separated with no overlap.
\label{f3}}
\end{figure}

A possible structure of the outflow is shown in Figure~\ref{f3}. The
$C_1$ absorber covers only the sightline toward image~A but not
image~B, regardless of its distance ($r$) from the flux source. The
size of the absorbing cloud ($d_{cloud}$) should be smaller than the
size of the flux source (because of partial coverage) and also smaller
than the physical distance between the sightlines of the lensed
images, i.e., $d_{cloud}$ $\leq$ $r \theta$ to avoid covering both
sightlines.  Here, we assume the separation angle of the two
sightlines from the flux source is very similar to that seen from us.
On the other hand, $C_2$ has two possible origins: a) small gas clouds
close to the flux source and b) filamentary (or sheet-like) structure
made of multiple clumpy gas clouds \citep{mis13}. In either case, both
sightlines A and B need to be covered.  We will discuss these further
in Section 4.3.

\subsection{Origin of Time Variation in \ion{C}{4} Lines}
Time variability is frequently detected in broad absorption lines
(BALs) with line widths of $\geq$ 2,000~\kms\ (e.g.,
\citealt{gib08,cap13,tre13}). Some narrower intrinsic absorption lines
(NALs and mini-BALs) are also known to be variable (e.g.,
\citealt{wis04,nar04,mis14}).  There are several explanations for the
time variation, including gas motion across our line of sight
\citep[e.g.,][]{ham08,gib08}, changes of the ionization condition
\citep[e.g.,][]{ham11,mis07b}, and a variable scattering material that
redirect photons around the gas clouds \citep[e.g.,][]{lam04}.  These
mechanisms are not applicable to intervening absorbers because they
have larger sizes (i.e., gas motion and photon redirection do not
work) and lower densities (i.e., the variability time scale due to ion
recombination is too long to observe over several years), compared to
intrinsic absorbers as noted in \citet{nar04}.

In our monitoring campaign, only \ion{C}{4} absorption lines show
clear time variation in both of the images. The absorption strength is
seen to weaken over the entire wavelength range (see Figure~\ref{f1}).
This immediately rejects the gas motion scenario because all gas
clouds that produce the $C_1$ and the $C_2$ need to cross our
sightline in concert, which is highly unlikely.  The scattering
scenario is also difficult to accept because it cannot explain the
fact that only \ion{C}{4} changes while \ion{N}{5} and \lya\ are
stable.  Thus, we conclude that the change of ionization scenario is
the most plausible explanation.

The $C_1$, arising in the absorber that locate only on the
sightline~A, was monitored twice in 2010 and 2014.  On the other hand,
the $C_2$ has two possibilities. If the corresponding absorber locates
on both sightlines toward the images~A and B, we have monitored the
variable \ion{C}{4} lines in four epochs, i.e., images~A and B in 2010
and then images~A and B in 2014, with time intervals of $\sim$ 744,
770, and 744~days, given the time delay of $\Delta t$ = 744~days
between images~A and B.  If the filamentary (or sheet-like) structure
is the case, it means we have monitored them only twice as we did for
the $C_1$.  In either case, we cannot monitor the ionization condition
of the absorbers because a wide range of ionic species (which is
necessary for photoionization modeling) are not detected in our
spectra.

Here, we discuss two possible scenarios for explaining the decrease of
the \ion{C}{4} line strength.  First, the ionization level may have
increased between the observations with more C$^{3+}$ ionized to
C$^{4+}$ while the ionization fraction of N$^{4+}$ remained stable. If
the absorber's ionization parameter\footnote[15]{The ionization
  parameter U is defined as the ratio of hydrogen ionizing photon
  density ($n_{\gamma}$) to the electron density ($n_e$), U $\sim$
  $n_{\gamma}/n_e$.} is $\log U$ $\sim$ $-$1.5, at which point the
ionization fraction of \ion{N}{5} is close to peak \citep{ham97a},
this scenario is possible\footnote[16]{For example, the ionization
  fraction of C$^{3+}$ changes by a factor of $\sim$2 while the
  corresponding change is only $\sim$5\%\ for N$^{4+}$ with a
  variation of $\Delta \log U$ $\sim$ 0.3 around $\log U$ $\sim$
  $-$1.5, assuming the continuum shape of typical quasar used in
  \citet{nar04}.}.  Alternatively, the invariance of \ion{N}{5} may be
due to the saturation effect with partial coverage.  Another possible
interpretation is recombination of C$^{3+}$ to C$^{2+}$.  In this
case, we can place constraints on the electron density and the
distance from the flux source by the same prescription as used in
\citet{nar04}, assuming the variation time scale as the upper limit of
the recombination time.  If we monitored the absorbers twice (i.e.,
$C_1$ and $C_2$ in the filamentary model) or four times (i.e., $C_2$
in the single-sightline model), the electron density is estimated to
be $n_e$ $\geq$ 8.7$\times$10$^{3}$~\cmmm\ or
1.72$\times$10$^{4}$~\cmmm, and the distance from the flux source to
be $r$ $\leq$ 620~pc or 440~pc, respectively. Because the absorber's
distance is always smaller than the boundary distance (see Section
4.3) in both cases, the filamentary model can be rejected for the
$C_2$ if recombination is the origin of the variation.

\subsection{Detectability of Sightline Difference}
Whether we detect sightline difference or not depends on the
absorber's size and its distance from the flux source.  For placing
constraints on the absorber's distance, the size estimation of the
background flux source is very important.  The outflow wind in
SDSS~J1029+2623 probably covers both the continuum source with a size
of $R_{\rm cont}$ $\sim$ 2.5$\times$10$^{-4}$~pc\footnote[17]{We
  assume $R_{cont}$ is five times the Schwarzschild radius, $R_{\rm
    cont}$ = 10GM$_{\rm BH}$/c$^2$.} and broad emission-line region
(BELR) with a size of $R_{\rm BELR}$ $\sim$ 0.09~pc\footnote[18]{This
  is calculated in \citet{mis13}, using the empirical relation between
  $R_{\rm BELR}$ and quasar luminosity \citep{kas00,mcl04}.} because
the residual flux at the bottom of the absorption lines are close to
zero around the peak of the broad emission lines (i.e., covering
factor toward BELR is \cf\ $\sim$ 1).  Following \citet{mis13}, we
define a boundary distance ($r_b$), a distance from the flux source
where the physical distance between two sightlines ($r_b \theta$) is
same as $R_{\rm BELR}$ (i.e., the two sight-lines become fully
separated with no overlap at $r$ $>$ $r_b$). If the BELR as well as
the continuum source is the background source, the boundary distance
is $r_b$ $\sim$788~pc\footnote[19]{If only the continuum source is the
  background source, which is applicable for absorption lines with
  large ejection velocity, the boundary distance is $r_b$
  $\sim$2.3~pc.}.
 
Here, we present two possible scenarios for the origin of the $C_1$.
First, the $C_1$ absorber could locate at larger distance than the
boundary distance, $\geq$788~pc. In this case, the $C_1$ absorber may
be the inter-stellar medium (ISM) of the host galaxy that are swept up
by an accretion disk wind \citep{kur09}. Another possible scenario is
that the absorber could be small clumpy cloud that locate at the
outskirt of the $C_2$ absorber whose distance is smaller than the
boundary distance.  \citet{ham13} suggests that mini-BAL absorbers
consist of a number of small gas clouds ($d_{cloud}$ $\leq$ 10$^{-3}$
-- 10$^{-4}$~pc) with very large gas density ($n_e$ $\geq$ 10$^6$ --
10$^7$ \cmmm) at the absorber's distance of $r$ $\sim$ 2~pc to avoid
over-ionization.  A similar picture is also suggested for BAL quasars
\citep{jos14}.  Furthermore, recent radiation-MHD simulations by
\citet{tak13} reproduce variable clumpy structures with typical sizes
of $\sim$ 20~$r_{\rm g}$ in warm absorbers, corresponding to $\sim$
5$\times$10$^{-4}$~pc assuming the black-hole mass of SDSS~J1029+2623,
M$_{\rm BH}$ $\sim$ 10$^{8.72}$ M$_{\odot}$.  Indeed, high-velocity
intrinsic NALs are frequently detected with partial coverage toward
the continuum source only, suggesting their typical size is comparable
to or smaller than the continuum source \citep{mis07a}.

\begin{figure}
 \begin{center}
  \includegraphics[width=12cm,angle=0]{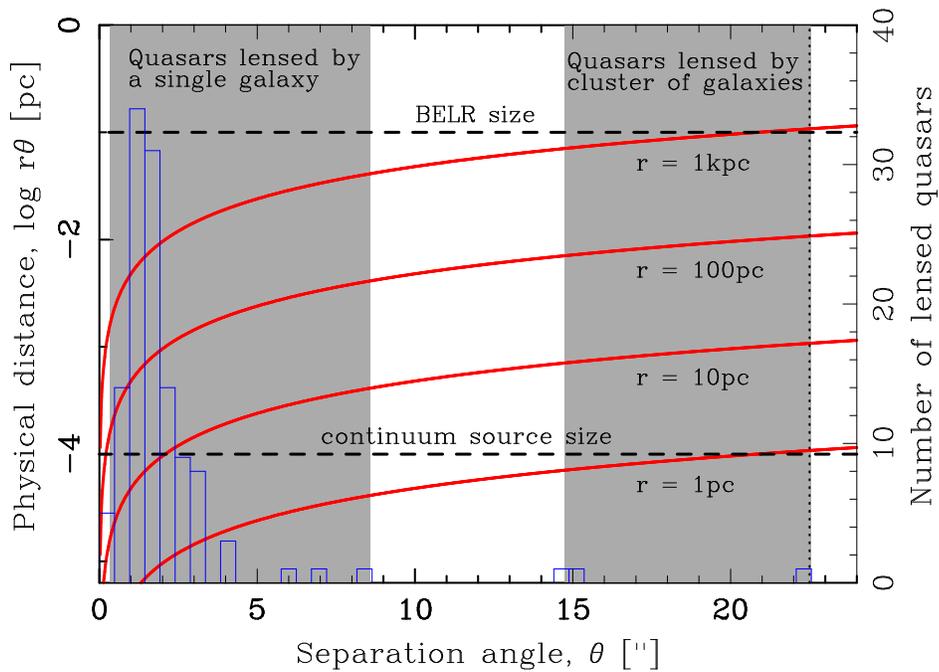}
 \end{center}
 \caption{Physical distance between two sightlines of lensed images as
   a function of separation angle (red curves), adopting the
   absorber's distance of $r$ = 1, 10, 100, and 1000~pc.  The
   horizontal dotted lines mark sizes of BELR ($\sim$0.1~pc) and the
   continuum source ($\sim$10$^{-4}$~pc) of SDSS~J1029+2623.  The
   histogram denotes the distribution of separation angles of 124
   lensed quasars that have been discovered to date.  Shaded regions
   are ranges of separation angles of quasar images lensed by a single
   galaxy or by a cluster of galaxies.\label{f4}}
\end{figure}

Our results have broader implications as well. Figure~\ref{f4}
summarizes physical distance between lensed images as a function of
separation angle for the 124 lensed quasars discovered to
date\footnote[20]{These are collected from the lensed quasar catalogs;
  CASTLE (http://www.cfa.harvard.edu/castles) and SQLS
  (http://www-utap.phys.s.u-tokyo.ac.jp/\~{}sdss/sqls).}, assuming the
absorber's distance is 1~pc, 10~pc, 100~pc, and 1~kpc.  The sightline
difference will be detected for quasars lensed by a single galaxy
whose typical separation angle is $\theta$ $\sim$ 2$^{\prime\prime}$,
if the absorber's size is smaller than its distance from the flux
source by more than five orders of magnitude (i.e., $d_{cloud}/r$
$\leq$ $\theta$). This would place an important constraint on the
absorbers.

\acknowledgments We thank the anonymous referee for a number of
comments that helped us improve the paper. We also thank Koji Kawabata
and Akito Tajitsu for their comments about data analysis.  The
research was supported by JGC-S Scholarship Foundation, and partially
supported by the Japan Society for the Promotion of Science through
Grant-in-Aid for Scientific Research 26800093.

\clearpage


\end{document}